# Cross-site Scripting Attacks on Android WebView

[1]Bhavani A B

[1]Hyderabad, Andhra Pradesh-500050, India

**Abstract**
WebView is an essential component in Android and iOS. It enables applications to display content from on-line resources. It simplifies task of performing a network request, parsing the data and rendering it. WebView uses a number of APIs which can interact with the web contents inside WebView. In the current paper, Cross-site scripting attacks or XSS attacks specific to Android WebView are discussed. Cross-site scripting (XSS) is a type of vulnerability commonly found in web applications. This vulnerability makes it possible for attackers to run malicious code into victim's WebView, through HttpClient APIs. Using this malicious code, the attackers can steal the victim's credentials, such as cookies. The access control policies (i.e., the same origin policy) employed by the browser to protect those credentials can be bypassed by exploiting the XSS vulnerability.

.
# 1. Introduction

Many of the Android applications display web content and also interact with it. This is possible by exposing a web browser as a standalone component and embedding it in the application. Such a component is called as WebView. It uses WebKit rendering engine to display web pages. It also enables developers to incorporate browser functionalities such as rendering, navigation etc. in the application.

1.1 Introduction to WebView APIs

There are two types of APIs in WebView, the Web-based APIs and the UI based APIs [2]. Web-based APIs are designed to interact with the web-contents inside the WebView. Examples of these APIs include loadURL, CookieManager.getCookie, etc. Attacks described in [1] target the Web-based APIs. WebView is subclass of a more generic View class.

View is the base class for widgets, which are used to create interactive UI components (buttons, text fields, etc). Therefore, WebView inherits the APIs of super class. Such APIs are UI-based APIs. Attacks describe in [2] target the UI based APIs. WebView APIs are described in detail in section 2.

1.2 Overview of the work and contribution

In the current paper, Cross-site scripting attacks on android WebView are investigated. Such attacks can be modeled using the interaction of Web-based APIs with the HttpClient. Using the should Override Urlloading() hook, the loadURL method can be intercepted, and the cookies can be stolen from the user's mobile. The scripts residing at the server can be run on the Android device and stolen cookies can be sent to the server through HttpGet and HttpPost APIs. The stealing of cookies may lead to several vulnerabilities such as Session Hijacking and impersonating the user through his cookies. The attacks may also result in stealing of sensitive information from the phone such as phone's contacts.

# 2. Web View APIs

The package which provides tools for Android application to browse the web is android.webkit. The package contains number of classes and interfaces. The most important class of the package is the WebView. It enables the developer to embed a built-in Web browser as widget, for displaying HTML content and browsing the web. In addition to WebView, android.webkit provides several other classes such as CookieManager, CookieSyncManager, WebChromeClient, WebViewClient etc. Jointly, these classes expose many APIs to Android applications. Based on their purposes, these APIs can be divided into two main categories, the Web-based APIs and UI based APIs [2]. In the current paper, we focus only on the Web-based APIs and their interaction with HTTP client. The paper also focuses on APIs related to two classes, WebViewClient and CookieManager.

2.1 Web-page related hooks

Android applications can monitor the events that occurred within WebView. This is done through the hooks provided by the WebViewClient class. Once triggered, these hook functions can access the event information, and may change the consequence of the events. To use these hooks, Android applications should first create a WebViewClient object, and then tell WebView to invoke the hooks in this object, when the intended events have occurred inside WebView. WebViewClient has already implemented the default behavior basically doing nothing for all the hooks. If we want to change that, we can override the hook functions with our own implementation. This is described in the code fragment below:

```
private class MyWebViewClient
extends WebViewClient { @Override
public boolean
shouldOverrideUrlLoading(WebView view,
          String url){
webView.loadUrl(url);
}
```

2.2 Cookie Manager

HTTP cookie, commonly referred to as just "cookie" is a parcel of text sent back and forth between a web browser and the server it accesses. Without a cookie, a web server cannot distinguish between different users, or determine any relationship between sequential page visits made by the same user. The CookieManager class in Android is used to manage cookies used by application's WebView. This class provides a number of public methods such as getCookie, acceptCookie() etc.

## 3. HttpClient APIs

HttpClient Class is an interface for HTTP client. The most essential function of HttpClient is to execute HTTP methods. Execution of an HTTP method involves one or several HTTP request / HTTP response exchanges, usually handled internally by HttpClient. The user is expected to provide a request object to execute and HttpClient is expected to transmit the request to the target server return a corresponding response object, or throw an exception if execution was unsuccessful.

```
HttpClient httpClient = new DefaultHttpClient();
```

As shown in the code above, DefaultHttpClient is the default implementation of the HttpClient interface.

3.1 HttpRequest

All Http Requests have a request line consisting of a method name, a request URI and a HTTP protocol version. HttpClient supports out of the box all HTTP methods GET, HEAD, POST, PUT, DELETE, TRACE and OPTIONS. There is a specific class for each method type: HttpGet, HttpHead, HttpPost, HttpPut, HttpDelete, HttpTrace, and HttpOptions. The Request-URI is a Uniform Resource identifier that identifies the resource upon which to apply the request. HTTP request URIs consists of a protocol scheme, host name, optional port, resource path, optional query, and optional fragment. The HttpPost request is show in the code fragment below:

```
HttpPost httpPost = new
HttpPost("http://maliciousScript/executeScript.php");
```

## 4. Attack Model

Following are the assumptions made in order to launch attacks successfully:

1. For the attacks described throughout the paper, the applications need to be granted with permission Android.permission.INTERNET. This permission is granted to 86.6% of free applications and 65 percent of paid applications [10]. Since this permission is granted to most of the applications, it is quite easy for the attacker to launch the attacks.
2. The attacks described in the case study in Section 3.2 needs to be granted with permission android.permission.READ_CONTACT. This permission is granted to 16.1% of free apps and 11% of paid apps [10]. Many of the free apps such as whatsApp are installed with this permission.
3. The paper describes the vulnerabilities in the third party Android applications. The owner of the web contents inside WebView, and the developer of the app are not the same. Therefore, there is a potential threat from a malicious application.

## 5. Cross-site scripting attacks

Cross-site scripting (XSS) is a type of computer security vulnerability typically found in Web Applications. To protect the user's environment from malicious code, browsers use a sand-boxing mechanism that limits a script to access only resources associated with its origin site. Unfortunately, these security mechanisms fail if a user unknowingly executes a malicious script from an intermediate, trusted site. In this case, the malicious script is granted full access to all resources (e.g. Authentication tokens and cookies) that belong to the trusted site. Such attacks are called cross-site scripting (XSS) attacks.

In the present work, Cross-site scripting attacks on Android WebView are investigated. By finding ways of executing malicious scripts through the third party malicious app on phone, an attacker can gain elevated access-privileges to sensitive page content, session cookies, and a variety of other information in the phone such as contacts. Cross-site scripting attacks in android are also investigated in [11]. The attacks described in [11] exploits flaw in the Intent handling mechanism of the Android browser. In the current paper, cross-site scripting attacks specific to WebView are investigated.

It has been pointed out by [9] that Android includes a number of mechanisms to reduce the scope of the potential issues by limiting the capability of WebView to the minimum functionality required by the application. addJavaScriptInterface() allows JavaScript to invoke operations that are normally reserved for Android applications. If un-trusted input is allowed, un-trusted JavaScript may be able to invoke Android methods. Leo et al. have modeled JavaScript Injection attacks using WebView's loadUrl() API[1]. The loadUrl() API receives an argument of string type; if the string starts with "javascript:", WebView will treat the entire string as JavaScript code, and execute it in the context of the web page that is currently displayed by the WebView component. This JavaScript code has the same privileges as that included in the web page. Essentially, the injected JavaScript code can manipulate the DOM tree and cookies of the page. In the present work, the possibility of launching a Cross-site Scripting attacks without using setJavascriptEnabled() API is investigated. The attacks can be modeled by executing the scripts residing at the server and sending the malicious content to server through Web-Service using HttpGet, HttpPost and HttpPut APIs. Such attacks result in cookie stealing, Session hijacking and impersonating user using stolen cookies. Each of these attacks is described in detail in the later sections.

5.1 Stealing cookies from the victim's device

The most common behavior of XSS attacks is to gather cookies. Cookies are small text files that reside on a user's computer and store name-value pairs along with some metadata. Cookies are commonly used to store information intended to be persistent during a browser session or from session to session, such as session IDs, user preferences, or login information. The cookie specifications assume that only the domain that set the cookie will be able to access it.

**Attack Method:** When the user runs the application through the WebView, Android applications can monitor the events occurred within WebView. We override the shouldOverrideUrlLoading hook, which is triggered by the navigation event, (when the user tries to navigate to another URL). Cookies can be gathered at every page navigation of the user using the method getCookie() from CookieManager class as shown in the code fragment below:

```
CookieManager cookieManager =
  CookieManager.getInstance();
final String cookie =
  cookieManager.getCookie(url);
```

Through HttpPost, malicious script can be run on the user's Android device, cookies and URL can be sent to any third party (i.e., the attacker's server), thus avoiding the same-origin policy or cookie protection mechanism.

```
HttpClient httpClient =
  new DefaultHttpClient();
HttpPost httpPost =
  new HttpPost("http://evilScript/androidCookie.php");
```

The attacker is now able get all the cookies and will be able to launch several attacks such a Session Hijacking and impersonating user using stolen cookies. The attacks described above are quite dangerous as the user sees only the trusted content and is not aware that his cookies are being stolen. The attack is shown in Figure 1.

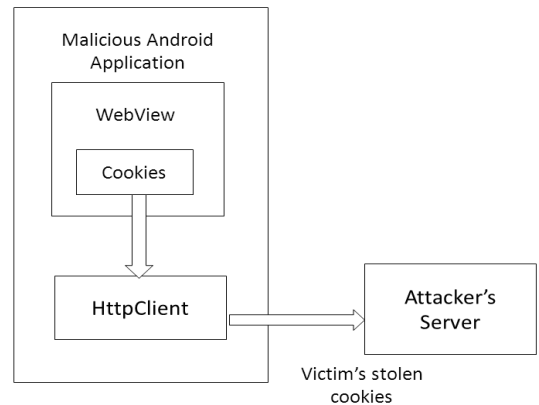

Fig 1: Cookie Stealing

**Session Hijacking:** Session hijacking is a method of taking over a Web user session by stealthily obtaining the session ID and gaining unauthorized access. The session ID is normally stored within a cookie or URL. Once the user's session ID has been accessed, the attacker can masquerade as that user and do anything the user is authorized to do on the network. From the web server's point of view, a request from an attacker has the same authentication as the victim's requests, thus the request is performed on behalf of the victim's session. This usually results in the attacker being able to perform all normal web application functions with the same privileges of legitimate user (e.g. online bill pay, composing an email, etc.).

**Impersonating user using stolen cookies:** If a website uses cookies as session identifiers, attackers can impersonate users after stealing victim's cookies. By stealing cookies of social networking sites, message boards or forums, the attacker can post a new message in the victim's name delete the victim's post or exploit user's credentials without his consent.

**Implementation:** To demonstrate the feasibility of the attacks, PHP script is used at the server side to process POST requests from the malicious application. The POST method transfers information through HTTP headers. $_POST associative array in PHP is used to access cookies sent from the Android device using POST. At the device side, the details of implementation are as follows:

1. Load the url of the target site using webView.loadUrl
2. Create WebViewClient object to invoke hooks in the object.
3. With shouldOverrideUrlLoading hook, the host application gets a chance to take over the control when a new URL is about to be loaded in the current WebView.
4. For each of the URL that is overloaded, cookies can be gathered using cookieManager.getCookie class.
5. The cookies can be sent to the attacker's server using HTTPclient POST method.

After the attacker gains access to the user's cookies, he can hijack the user's sessions or impersonate the user using the cookies. The feasibility of such attacks were verified using the open source PHP based message board called phorum[12]. When the user has logged into the message board through malicious application, using his username and password, the cookies from the device are sent through POST method to the attacker's server. The attacker can use the stolen cookies and login to the message board with victim's credentials. The attacker can access the data in victim's account and also impersonate the victim by posting content on victim's name.

3.2 Accessing Sensitive information from the Android Phone

An XSS attack could also use vulnerabilities in android WebView to scrape useful information out of phone such as phone's contacts, email ids and phone numbers. The user is completely unaware of such attacks as the user views the content from trusted web-pages.

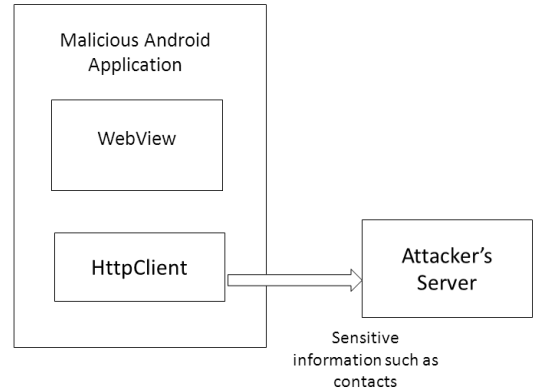

Fig 2: Stealing sensitive information from phone

**Case Study:** In the following case study, a Cross-site scripting attack in which phone's contacts are sent to the attacker's server is demonstrated. Let us assume that the user installs Facebook app, which is one of the most popular third party app. The user views the Facebook login page in the WebView. The malicious application fetches the sensitive information from the phone such as user's contact details, phone numbers, Email-id etc., and sends it to the attacker's server through the HttpClient as shown in Figure 2. The attack is very easy to launch but difficult to detect. The user is not aware of such attack as he views only the legitimate content.

**Implementation:** To demonstrate the feasibility of the attack, a sample malicious Android application was developed. The details of the application are as follows:
1. Load the url of the target site using webView.loadUrl.
2. Sensitive information such as user's contacts can be sent to the attacker's server using HTTPclient POST method. User views only the legitimate content in his WebView and does not know that sensitive information is stolen.

PHP scripts are used at the server side to process POST requests from the malicious application.

## 6. Conclusion

In the present work, Cross-site scripting attacks with respect to Android WebView and HttpClient are studied. Such type of attack results in stealing of cookies and other sensitive information such as contacts from the Android phone. They also result in Session Hijacking and impersonating user using stolen cookies. The attacks are a result of breach in the same origin policy of Android browsers. XSS attacks are easy to execute, but difficult to detect and prevent. The future work will focus on building solutions to defend against the attacks on WebView.

**First Author** Bhavani A B has done her M.Tech in VLSI and Embedded Systems from International Institute of Information Technology, Hyderabad. She has over 3 years of experience in the Mobile industry in areas related to Embedded Systems, Mobile technologies and WebKit browser development. Her areas of interest include Embedded Systems, Mobile technologies, Mobile Security, WebKit, Digital Signal processing, audio and video codec.